\documentclass[aps,pra,twocolumn,showpacs,superscriptaddress,amsmath,amsfonts,amssymb,floatfix,nofootinbib]{revtex4-1}
\usepackage{graphicx}
\usepackage{dcolumn}
\usepackage{bbm}

\newcommand{\be}{\begin{equation}}
\newcommand{\ee}{\end{equation}}

\begin{document}

\title{Quantum coherence and sensitivity of avian magnetoreception}

\author{Jayendra N. Bandyopadhyay}
\email{jnbandyo@gmail.com}
\affiliation{Centre for Quantum Technologies, National University of Singapore, 3 Science Drive 2, 117543 Singapore}

\author{Tomasz Paterek}
\affiliation{Centre for Quantum Technologies, National University of Singapore, 3 Science Drive 2, 117543 Singapore}

\author{Dagomir Kaszlikowski}
\affiliation{Centre for Quantum Technologies, National University of Singapore, 3 Science Drive 2, 117543 Singapore}
\affiliation{Department of Physics, National University of Singapore, 2 Science Drive 3, 117542 Singapore}

\begin{abstract}
Migratory birds and other species have the ability to navigate by sensing the geomagnetic field.
Recent experiments indicate that the essential process in the navigation takes place in bird's eye and uses chemical reaction involving molecular ions with unpaired electron spins (radical pair).
Sensing is achieved via geomagnetic-dependent dynamics of the spins of the unpaired electrons.
Here we utilize the results of two behavioral experiments conducted on European Robins to argue that the average life-time of the radical pair is of the order of a microsecond
and therefore agrees with experimental estimations of this parameter for cryptochrome --- a pigment believed to form the radical pairs.
We also found a reasonable parameter regime where sensitivity of the avian compass is enhanced by environmental noise,
showing that long coherence time is not required for navigation and may even spoil it.
\end{abstract}

\date{\today}
\maketitle

Recently there has been a growing interest in the application of quantum mechanics to understand many biological phenomena such as photosynthesis \cite{PhotoEngel,PhotoEnt2010,PhotoCoh2007,CohMigrRoom2009,PH2008,PhotoQuantumWalk,ExcitationTransfer_NoiseAssistedTrans}, process of olfaction \cite{OLF2002, ODOR2007}, enzymatic reactions \cite{AmoniaRadical, Enz} or avian magnetoreception \cite{Ritz2000,Ritz2009,Cai_PRL2010,VlatkoPRL}. These interests have brought physicists, chemists, and biologists at the same platform and led to the beginning of a new interdisciplinary subject called {\it quantum biology} \cite{AJV2009,BALL}. A major motivation of these studies is to understand how nature utilizes purely quantum phenomena to optimize various biological processes. 

Here we are specifically interested in the avian magnetoreception.
It is very plausible that the navigation ability of some migratory birds is governed by the mechanism based on geomagnetic-dependent dynamics of spins of unpaired electrons in a radical pair.
A recent theoretical study has estimated both the life-time of the pair and the coherence time of this dynamics to be of the order of tens of microsecond \cite{VlatkoPRL}.
The basic criterion used there postulates that bird's navigation is disturbed if the signal produced by the dynamics is independent of the orientation of the geomagnetic field.
This criterion together with the results of behavioral experiments in which European Robins could not navigate in a weak oscillating magnetic field \cite{RES_RAD_PAIR_COMPASS, RF_NATUR} led to the estimated life time and coherence time.
Here we additionally take into account the results of other behavioral experiments in which the same species were observed to be temporarily disoriented in a constant magnetic field sufficiently stronger or weaker than the geomagnetic field \cite{FuncWindowBook, FuncWindow}.
We estimate the life time and coherence time of the order of several microseconds. Our estimate is consistent with that obtained in a recent behavioral experiment \cite{Ritz2009} and also with the {\it in vitro} experiment using cryptochrome \cite{Biskup09}, a pigment believed to form the radical pairs.
Furthermore, we demonstrate theoretically that environmental noise can enhance the sensitivity of the avian compass,
i.e. sensitivity in the presence of noise is better than without noise.
This increase of sensitivity has resonant character and shows that long coherence times sometime may be disadvantageous for navigation.

Let us begin with certain functional properties of avian compass which have been observed in different behavioral tests:
(i) A very early experiment with European Robin showed that in contrast to the well-known magnetic material based physical compass the avian compass does not depend on the polarity of magnetic field but only on the inclination of magnetic field \cite{Frankfurt1972};
(ii) An experiment in which geomagnetic field was supplemented with a very weak radio-frequency (RF) field showed that birds disoriented at the frequency of the RF field which is resonant with the energetic splitting of a free electron (due to its spin) induced by the local geomagnetic field. 
This was observed only when the RF field was not parallel to the local geomagnetic field \cite{RES_RAD_PAIR_COMPASS, RF_NATUR};
(iii) Another experiment showed that the avian compass works within a narrow ``functional window'' around the local geomagnetic field. The compass ceased functioning when intensity of the magnetic field increased/decreased by about $30\%$ of the local geomagnetic field. However, birds exposed to the new intensity of the magnetic field sufficiently long adapted themselves and their compass again worked correctly \cite{FuncWindowBook, FuncWindow}.

These experimental findings are consistent with the following most widely used model of avian magnetoreception (see Ref. \cite{RodgersPNAS} for a recent review). In Ref. \cite{Erik-Simon}, a new type of RP based model has very recently been proposed.
It is assumed that bird's retina contains a photoreceptor pigment with molecular axis direction dependent on its position in the retina.
Absorption of incident light by a part of the pigment results in electron transfer to a suitable nearby part and in this way a radical pair is formed, i.e. a pair of charged molecules each having an electron with unpaired spin. 
In the external magnetic field the state of electron spins undergoes singlet-triplet transitions and at random times the radical pairs recombine forming singlet (triplet) chemical reaction products. 
The amount of these chemical products varies along the retina as the direction of molecular axis changes, and the shape of this profile is believed to be correlated in bird's brain with the orientation of the geomagnetic field.

The mathematical model of this mechanism of a chemical reaction based avian compass involves coupling of electron spins to external magnetic field and to the spins of the molecular core.
The property (ii) suggests that the spin of one of the electrons in the radical pair is effectively uncoupled from spins of any other particles and only interacts with the external field.
In a simple qualitative model the spin of the other electron is coupled to effective spin-$\frac{1}{2}$ core of the molecular part (``nuclear'' part)~\cite{VlatkoPRL}.
The corresponding Hamiltonian is as follows:
\be
H = \widehat{I}\,\centerdot\, \overleftrightarrow{\mathbf{A}}\,\centerdot\, \widehat{S}_1 + \gamma\, \mathbf{\overrightarrow{B}} \,\centerdot\, (\widehat{S}_1 + \widehat{S}_2),
\label{hamiltonian}
\ee
where $\widehat{I}$ is the nuclear spin operator, $\overleftrightarrow{\mathbf{A}}$ is the hyperfine (HF) tensor, $\widehat{S}_i \equiv \{\sigma_x^{(i)},\, \sigma_y^{(i)},\, \sigma_z^{(i)}\}$ are the electronic spin Pauli operators, $\mathbf{\overrightarrow{B}}$ is the magnetic field, and $\gamma = \frac{1}{2} \mu_0 g$ is the gyromagnetic ratio with $\mu_0$ being Bohr magneton and $g$ the electronic $g$-factor. Here we assume that the $g$-factors are the same for both electrons and set it at the value corresponding to the free electron, i.e. $g=2$. Following Ref.~\cite{VlatkoPRL} the HF tensor is chosen axial and anisotropic, i.e., $\overleftrightarrow{\mathbf{A}} = \mbox{diag}\{A_x,\,A_y,\,A_z\}$ with $A_x = A_y = a = A_z/2$. We define a quantity $A \equiv (A_x^2 + A_y^2 + A_z^2)^{1/2}$ as a measure of HF coupling strength and $\overline{a} = h/A$ as a measure of this strength in the time units, $h$ stands for Planck constant. Here $\overline{a} = h/\sqrt{6} a$. 

In order to verify how this model recovers the properties of behavioral experiments, one varies external magnetic field
\be
\begin{split}
\mathbf{\overrightarrow{B}} &= B_0\,(\sin \theta\,\cos \phi,\,\sin \theta\,\sin \phi,\,\cos \theta)\\  
&+ B_{\mbox{\small rf}}\,\cos(\omega t)\,(\sin \alpha\,\cos \beta,\,\sin \alpha\,\sin \beta,\,\cos \alpha),
\end{split}
\ee
where $B_0$ gives the strength of the local geomagnetic field or the artificially changed constant field, 
and $B_{\mbox{\small rf}}$ is the additional RF-field with frequency $\omega$, which is switched on optionally. For all our numerics, the RF field is orthogonal to the static field.
Due to the axial symmetry of HF tensor we set $\phi = 0$ without loss of generality.
We also assume that $\beta = 0$ for the oscillating field's direction.
Since most of the behavioral experiments were performed in Frankfurt, we set the magnetic field strength $|\overrightarrow{B_0}| \equiv B_0 = 47\,\mu T$ as the reference value.
The corresponding Larmor precession period is $h / 2 \gamma B_0 = 0.76\,\mu s.\,$\footnote{The factor of two emerging here is due to our definition of $\widehat{S}_i$ as Pauli operators.}
The HF interaction strength $\overline{a}$ typically acquires values in the range $10\,n s - 1 \, \mu s$~\cite{RodgersPNAS},
but we have verified that only for those radicals with the coupling strength close to the corresponding geomagnetic strength of $0.76\,\mu s$ present model predicts working compass.
Therefore, here we study two different cases of $\overline{a} = 1.0\, \mu s$ and $0.5\, \mu s$, which are slightly greater and smaller than the geomagnetic strength.

The dynamics of the system consisting of two electrons and one nuclei depends on two processes: coherent evolution determined by the above Hamiltonian and recombination of the pair to create spin-state dependent chemical products. The spin-chemistry community models the above processes following the Haberkorn approach \cite{haberkorn} according to which the dynamics of the density matrix $\rho$ describing the whole system is governed by equation
\be
\frac{d\rho}{dt} = - i [H,\,\rho] - \frac{k}{2}\,(Q_{\mbox{\tiny S}} \rho + \rho Q_{\mbox{\tiny S}}) - \frac{k}{2}\, (Q_{\mbox{\tiny T}} \rho + \rho Q_{\mbox{\tiny T}}),
\label{haberkorn} 
\ee
where $k$ determines the reaction rates for the singlet and triplet recombination, here assumed to be the same, and $Q_{\mbox{\tiny S}}$ and $Q_{\mbox{\tiny T}}$ are the projection operators onto the singlet and triplet subspaces.
Although other treatments are possible~\cite{Jones-Hore2010,Jones-Hore2011}, here we follow the Haberkorn approach as it is the most consistent for the spin-selective recombination of radicals \cite{Ivanov2010}.
We begin the evolution at the moment of radical pair creation with the initial density matrix
\be
\rho(0) = \frac{1}{2} \mathbbm{1}_N \otimes |\psi^-\rangle \langle \psi^-|,
\ee
where the electron pair is in the singlet state $|\psi^-\rangle = \frac{1}{\sqrt{2}}(|\uparrow \downarrow \rangle - |\downarrow \uparrow \rangle)$, and the nuclear spin is in a completely mixed state $\frac{1}{2} \mathbbm{1}_N$. 
The singlet product yield is defined as the amount of product decaying via the singlet channel
\be
\Phi_S = k \int_0^\infty \langle \psi^- | \mbox{Tr}_N \rho(t) |\psi^- \rangle \,dt,
\ee
where the partial trace of the density matrix is taken over the nucleus subspace.

We proceed to estimate the parameter $k$, inverse of which gives the average life-time of the radical pair.
For this purpose, we use behavioral tests (ii) and (iii) which describe bird's disorientation under the influence of RF field and weaker artificial geomagnetic field.
We shall not be interested here in the process of bird adaptation to the artificial magnetic field and only use the fact that the latter disturbs the compass.
Nevertheless we would like to emphasize that adaptation is a physiological process, therefore its explanation does not warrant any modifications of the present model.  
The model describes contribution to the chemical products from a single radical pair whereas the brain receives a cumulated signal from all created pairs (that are assumed to evolve independently).
Under the change of magnetic field intensity the cells of the retina may start to vary the amount of created pairs and in this way control the intensity of the cumulated signal.
Similar mechanism is present in vertebrate photoreceptors that can adapt their pigment sensitivity to external light by varying concentration of calcium ions \cite{ADAPT_REV}.
As an experimental indication that these two adaptation phenomena may share common mechanism we note that the full avian compass adaptation to external magnetic field intensity takes about the same time as adaptation of vertebrate photoreceptors to external light intensity.

\begin{figure}[t]
\centering
\includegraphics[width=8.5cm,height=6.5cm]{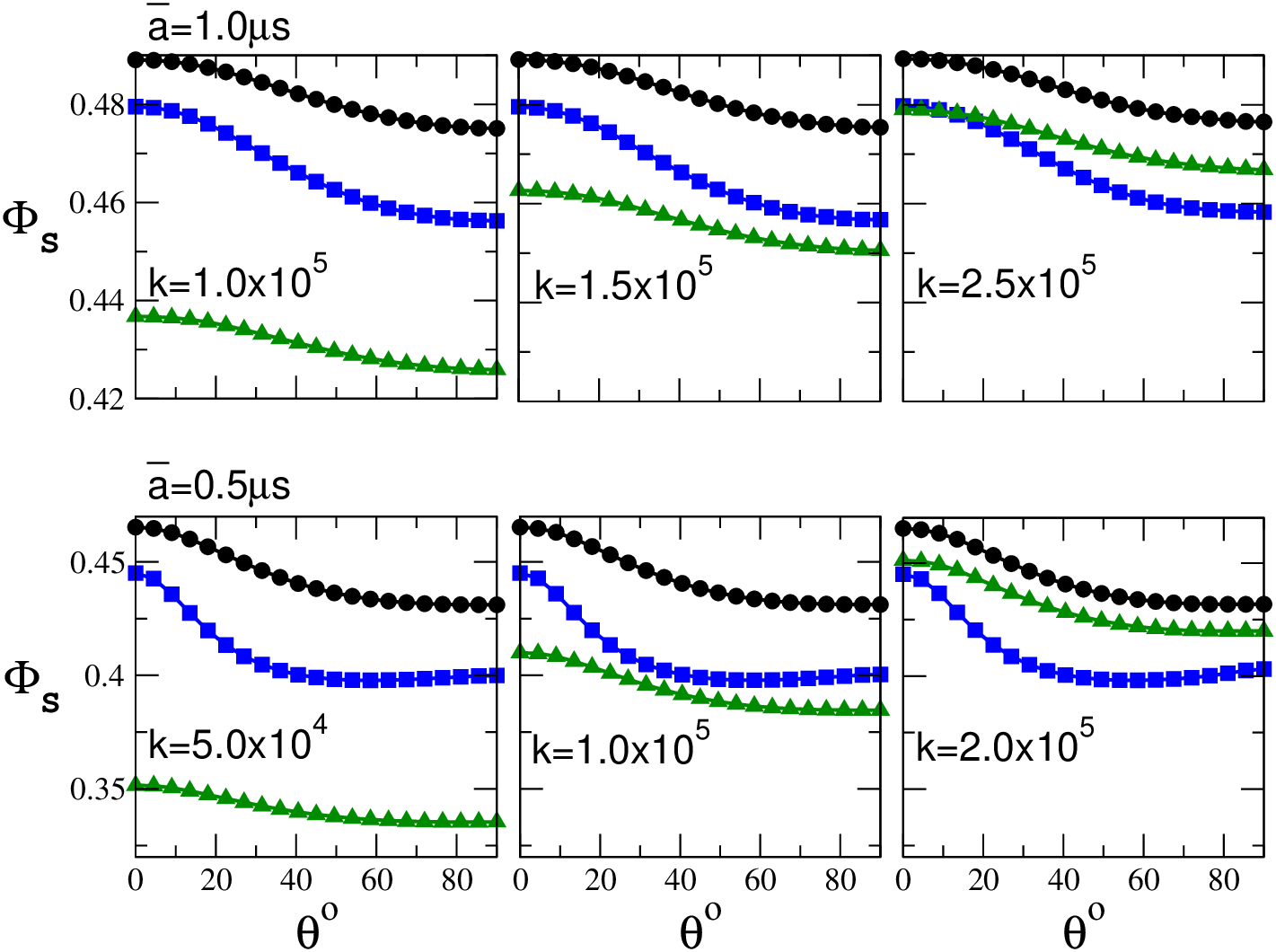}
\caption{(Color online) Life-time estimation of a radical pair.
Each plot presents angular dependency of the singlet yield in presence of local geomagnetic field of $47\,\mu T$ $(\bullet)$, in $30\%$ weaker field $(\blacksquare)$, and for the geomagnetic field augmented with the resonant RF field $(\blacktriangle)$.
Calculations are performed for two strengths of the hyperfine coupling: slightly smaller than the geomagnetic strength ($\overline{a} = 0.5\, \mu s$, three lower plots) and slight bigger than the geomagnetic strength ($\overline{a} = 1.0\, \mu s$, three upper plots).
On plots from left to right we display the growing value of the recombination rate $k$ used in numerics. 
The radical pair life-time is the inverse of $k$.
Since birds are disoriented both in the weaker field and in the presence of the RF field we postulate that
the criterion for estimating $k$ should be based on the comparison of the profile of triangles to the profile of squares.
This leads to the average life-time of the order of several microseconds (see main text for details)
in agreement with experiments on pigment believed to be present in the bird's retina.}
\label{ang_dep}
\end{figure}

Fig.~\ref{ang_dep} presents the angular dependence of the singlet yield for different $k$ values and for three different kinds of magnetic field environment: 
in the geomagnetic field of $47 \mu T$, in $30\%$ weaker magnetic field as in the experiments~\cite{FuncWindowBook, FuncWindow}, 
and in the geomagnetic field augmented by weak resonant RF field of $B_{\mbox{\small rf}} = 150 nT$ as in the experiment \cite{Ritz2009}, and also for a fair comparison with Ref. \cite{VlatkoPRL}.
All the plots show that the singlet yield angular profile changes slightly when the magnetic field intensity is reduced by $30\%$.
Note that there is a clear angular dependency of the profile both in the original and the weaker field.
Since the behavioral tests showed that the weaker magnetic field disrupts the avian compass, 
the degree of flatness of the angular profile cannot be the only property responsible for avian navigation.
The profile has to rather be within a proper range from the profile corresponding to the geomagnetic field alone.
Therefore when the resonant RF is added to the geomagnetic field, it can disrupt the avian compass not only by completely washing out the angular dependency of the profile \cite{VlatkoPRL} but also by changing the amount of singlet yield such that the profile is outside the range estimated by applying the weaker magnetic field. 
We use this criterion to estimate the parameter $k$.
Consider first three upper plots of Fig. \ref{ang_dep} that present results of calculations for the HF strength of $\overline{a} = 1.0\, \mu s$.
The left plot assumes $k = 10^5$ and shows that the profile corresponding to the RF field is much below the profile corresponding to the weaker magnetic field.
Similar behavior is observed for lower values of $k$ which we have not presented here.
In the middle plot ($k = 1.5 \times 10^5$) the RF field profile is slightly below the profile corresponding to the weaker field and therefore correctly predicts that bird's navigation is disturbed.
The right plot of $k = 2.5 \times 10^5$ shows to the contrary that the RF field profile is slightly above the profile corresponding to the weaker magnetic field, and therefore avian compass is not disturbed in contradiction with experiments.
This implies that all values of $k$ below $2 \times 10^5$ are consistent with experiments. 
We use the critical value $k = 2 \times 10^5$ for further calculations.
The corresponding average life-time of the radical pair is $5 \mu s$.
Similar analysis applied to the three lower plots, for the case of $\overline{a} = 0.5\, \mu s$, gives estimation of $k = 1.5 \times 10^5$ and the corresponding average life-time $6.7\, \mu s$.
These estimations are come close to the experimental estimations for the cryptochrome \cite{Biskup09}, which is believed to be the photo-receptor pigment responsible for the radical pair based avian compass of European Robins; and they also come within the range $2 \mu s - 10 \mu s$, the lifetime estimated in a recent behavioral experiment \cite{Ritz2009}. We have also done the same analysis using stronger $(B_{\mbox{\small rf}} = 470 nT)$ and weaker ($B_{\mbox{\small rf}} = 47 nT$) resonant RF-field, which are respectively $1\%$ and $0.1\%$ of the local geomagnetic field strength, and both have been reported to disturb avian compass \cite{Ritz2009}. For these two cases, we estimate the life-time about $1-2 \mu s$ and $11.0-12.5 \mu s$, respectively \cite{supplementary}. These estimated lifetimes are within or very close to the above mentioned range. We also study the case where the geomagnetic field $(47 \mu T)$ is replaced by the artificially created stronger ($94 \mu T$) static field. The estimated life-time for this case is also within the above mentioned range \cite{supplementary}. 

After setting the value of decay parameter $k$, we investigate the effect of environment on the singlet yield. 
Following Ref. \cite{VlatkoPRL} we describe the environment by a standard Lindblad formalism \cite{LINDBLAD1976,GKS1976}:
\begin{eqnarray}
\label{lindblad_eq}
\frac{d\rho}{dt} = - i [H,\,\rho]  - \frac{k}{2}\,(Q_{\mbox{\tiny S}} \rho + \rho Q_{\mbox{\tiny S}}) - \frac{k}{2}\, (Q_{\mbox{\tiny T}} \rho + \rho Q_{\mbox{\tiny T}})\nonumber\\
+ \sum_{i} \Gamma_i  \left[ L_i \rho L_i^\dagger - \frac{1}{2}\left(L_i^\dagger L_i \rho + \rho L_i^\dagger L_i \right) \right],~~~~~~~~~~~
\end{eqnarray}
where there are six noise operators $L_i$ altogether, three acting on the first electron and three acting on the second electron.
The three noise operators for each electron are taken to be the Pauli matrices $\sigma_x$, $\sigma_y$ and $\sigma_z$.
We assume that the noise parameters $\Gamma_i$ are all equal and given by the noise level $\Gamma$.
Our aim is to study influence of the noise level on the magnetic sensitivity of the compass, defined as~\cite{Cai2012}:
\be
\Delta_{\mbox{\bf\small S}} \equiv \Phi_S^{\max} - \Phi_S^{\min},
\label{SENSITIVITY}
\ee
where $\Phi_S^{\max}$ ($\Phi_S^{\min}$) is the maximum (minimum) singlet yield optimized over angle $\theta$.

\begin{figure}[t]
\centering
\includegraphics[width=8.5cm,height=7cm]{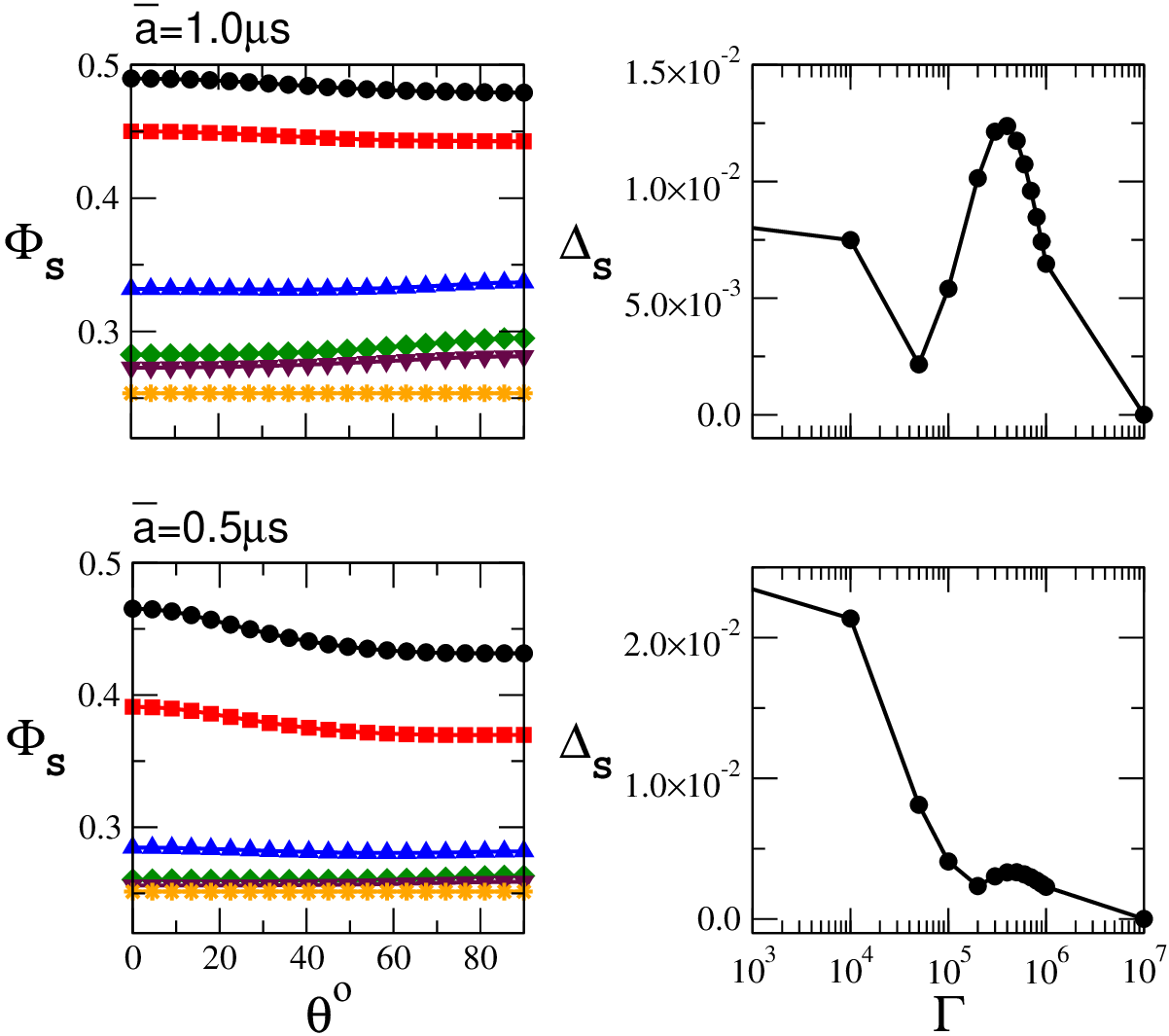}
\caption{(Color online) Magnetic sensitivity of avian compass in presence of environmental noise.
The upper plots present results of calculations for hyperfine coupling strength of $\overline{a} = 1\,\mu s$ and $k=2 \times 10^5$, the lower plots are for $\overline{a} = 0.5\,\mu s$ and $k=1.5 \times 10^5$.
The left plots present angular dependence of the singlet yield for different environmental noise levels [$\Gamma=0\,  (\bullet),\Gamma = 10^4 \, (\blacksquare), \Gamma = 10^5 \, (\blacktriangle), \Gamma = 4 \times 10^5 \, (\blacklozenge), \Gamma = 8 \times 10^5 \, (\blacktriangledown)$, and $\Gamma = 10^7 \, (*)$].
On the right plots we present the magnetic sensitivity defined in Eq. (\ref{SENSITIVITY}).
They reveal resonant character of the sensitivity for noise levels close to $\Gamma = 4 \times 10^5$, which correspond to optimal decoherence time of the order of a microsecond.
Moreover, the upper right plot shows that the sensitivity is increased in the presence of environment.
}
\label{decohere}
\end{figure}

Fig. \ref{decohere} presents results of calculations of the singlet yield and the magnetic sensitivity for different environmental noise levels.
For noise levels below $\Gamma \approx 10^4$ we find that the sensitivity is practically independent of environment.
Therefore, for decoherence time $T_D \equiv \Gamma^{-1} \approx 100 \, \mu s$ and longer, the sensitivity is practically the same as if there is no environment.
For shorter decoherence times (bigger $\Gamma$) the noise does influence the sensitivity and for the HF strength of $\overline{a} = 1.0\,\mu s$ we find quite counterintuitively that the sensitivity in the presence of noise is better than without noise ($\Gamma = 0$).
This is not so for $\overline{a} = 0.5\,\mu s$, but in both cases the sensitivity is displaying resonant behavior as a function of $\Gamma$ with its local maximum at the noise level $\Gamma = 4 \times 10^5$, see supplementary material \cite{supplementary} for further evidence that this resonant noise level does not depend on $k$. 
Therefore the best magnetic sensitivity for $\overline{a} = 1.0\, \mu s$ case takes place when the decoherence time $T_D \equiv \Gamma^{-1} = 2.5\, \mu s$. Here longer decoherence time is not useful as it can spoil the sensitivity. 

In summary, we have used the results of two different behavioral tests preformed with European Robins to estimate the average life time of the radical pair taking part in the avian magnetoreception. Unlike a recent study which took into consideration the result of only one behavioral test and estimated the average life time close to $100\,\mu s$ \cite{VlatkoPRL}, our estimation of the life time is about few microseconds which agrees well with experiments. As the most important result of the present work we consider identification of a parameter regime where the presence of environment enhances performance of the chemical compass.
Similar enhancement caused by environment is also found in studies of energy transfer during photosynthesis \cite{PHOTO_NOISE_PLENIO,PhotoQuantumWalk}, and very recently in avian magnetoreception \cite{Cai2012,Briegel-2012}. These suggest that Nature might be optimizing performance of some biological processes by utilizing inevitable noise present in environment.
More insight into this conjecture can be obtained from further studies of identified here resonance that magnetic sensitivity displays as a function of environmental noise.

\begin{acknowledgments}

This research is supported by the National Research Foundation and Ministry of Education in Singapore. We thank Erik Gauger and Simon C. Benjamin for helpful discussion. JNB thanks Bijay Kumar Agarwalla for extensive interaction. 

\end{acknowledgments}

\begin{appendix}

\section{}

In the main text, we use two behavioral experiments to estimate the life-time and the coherence time of the radical pair (RP). We set the static magnetic field at $47 \mu T$, which is the local geomagnetic field at Frankfurt, the site of all behavioral experiments. The life-time of RP is estimated using resonant RF field with strength $150 nT$, the RF field which was used in a recent behavioral experiment. In the same experiment, birds' behavior was also studied for other  RF field strengths and also for stronger static magnetic field. This supplementary material (SM) discusses results which we obtain using some of those different parameter values. They show that our estimations of the time scales are robust and do not depend much on the strength of RF field, and our identification of the resonance of magnetic sensitivity does not depend much on the life time of the radical pair.

\subsection{Life-time estimation for RF field of $470 nT$}

\begin{figure}[t]
\centering
\includegraphics[width=8cm,height=6cm]{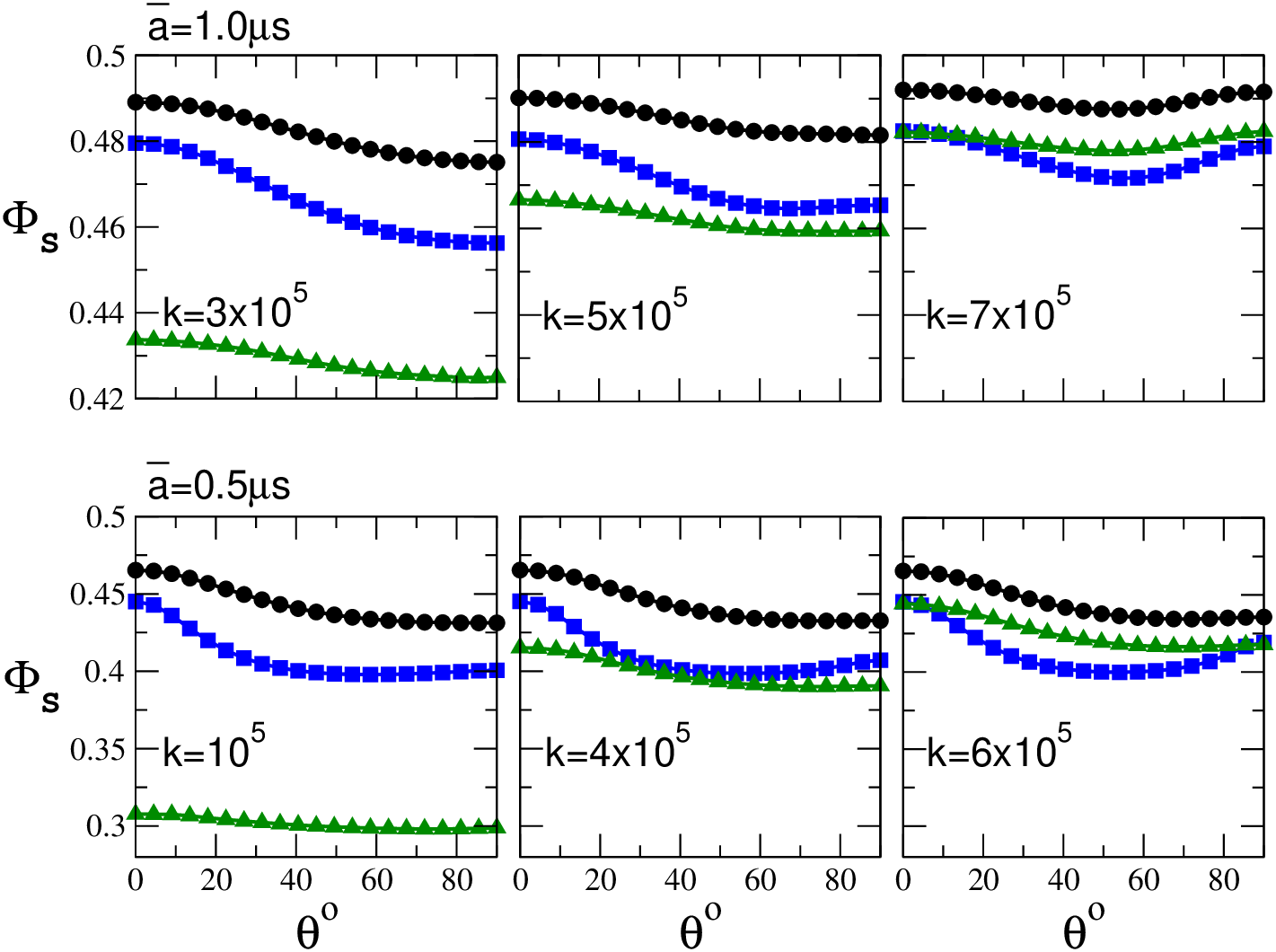}
\caption{(Color online) Life-time estimation of a radical pair using $470nT$ RF field.
Each plot presents angular dependency of the singlet yield in presence of local geomagnetic field of $47\,\mu T$ $(\bullet)$, in $30\%$ weaker field $(\blacksquare)$, and for the geomagnetic field augmented with the resonant $470 nT$ RF field $(\blacktriangle)$.
Calculations are performed for two strengths of the hyperfine coupling: slightly bigger than the geomagnetic strength ($\overline{a} = 1.0\, \mu s$, three lower plots) and slight smaller than the geomagnetic strength ($\overline{a} = 1.0\, \mu s$, three upper plots).
On plots from left to right we display the growing value of the recombination rate $k$ used in numerics. 
The radical pair life-time is the inverse of $k$, and can be estimated using the argument of the main text.}
\label{ang_dep1}
\end{figure}

In Fig. \ref{ang_dep1}, we show the angular dependence of the singlet yield for different decay parameter $(k)$ values \cite{Ritz2009} and for three different kinds of magnetic field environment: in the geomagnetic field of $47 \mu T$, in $30\%$ weaker magnetic field, and in the geomagnetic field augmented by weak resonant RF field of $B_{\mbox{\small rf}} = 470 nT$ ($1\%$ strength of the local geomagnetic field). Applying the argument of the main text gives estimation of the life-time $1.67 \mu s$  for $\overline{a} = 1 \mu s$, and $2 \mu s$ for $\overline{a} = 0.5 \mu s$.

\begin{figure}[ht]
\centering
\includegraphics[width=8cm,height=6cm]{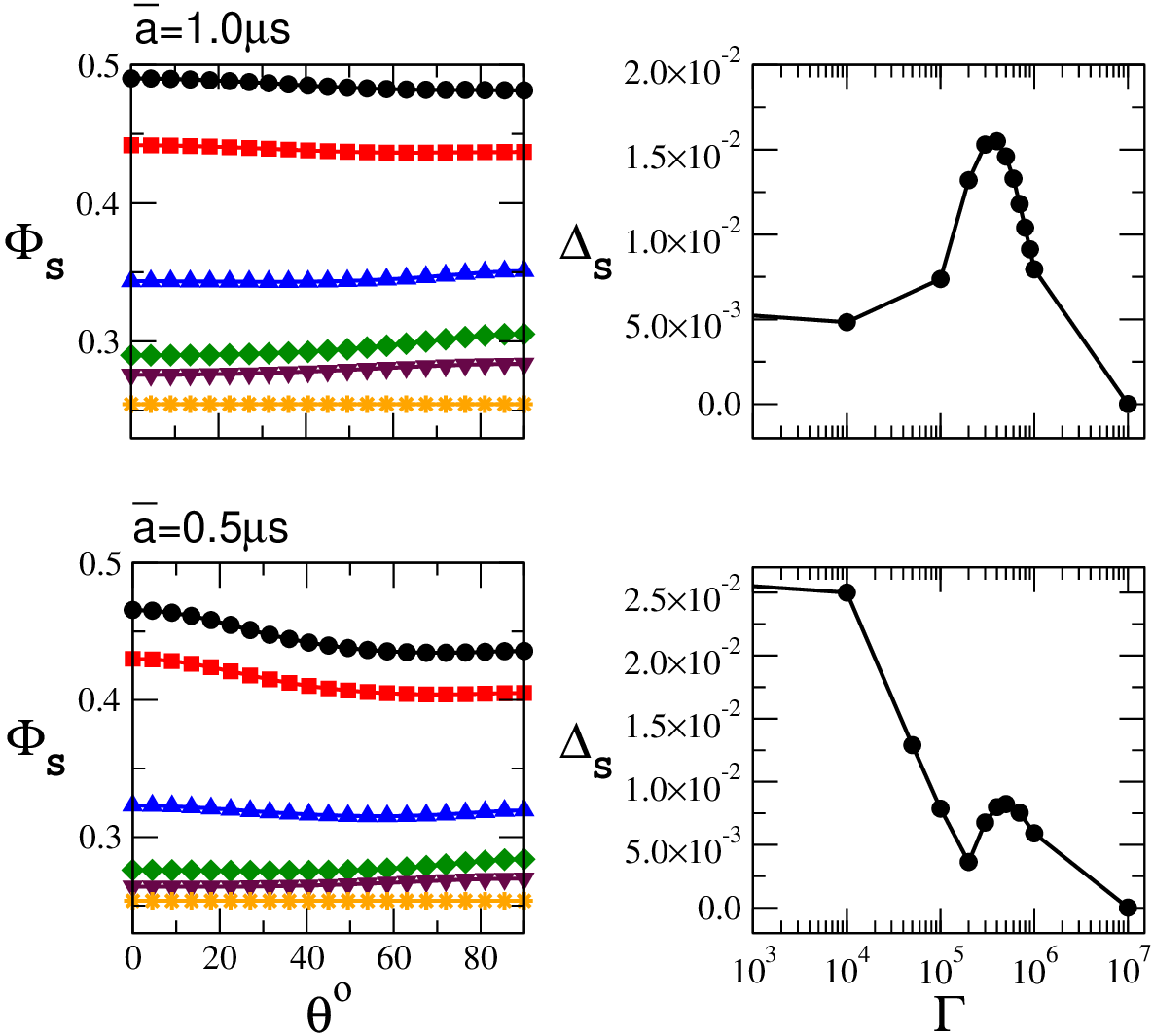}
\caption{(Color online) Magnetic sensitivity of avian compass in presence of environmental noise.
The upper plots present results of calculations for hyperfine coupling strength of $\overline{a} = 1\,\mu s$ and $k=2 \times 10^5$, the lower plots are for $\overline{a} = 0.5\,\mu s$ and $k=1.5 \times 10^5$.
The left plots present angular dependence of the singlet yield for different environmental noise levels [$\Gamma=0\,  (\bullet),\Gamma = 10^4 \, (\blacksquare), \Gamma = 10^5 \, (\blacktriangle), \Gamma = 4 \times 10^5 \, (\blacklozenge), \Gamma = 8 \times 10^5 \, (\blacktriangledown)$, and $\Gamma = 10^7 \, (*)$].
On the right plots we present the magnetic sensitivity defined in Eq. (7) of the main text. They reveal resonant character of the sensitivity for noise levels close to $\Gamma = 4 \times 10^5$, which correspond to optimal decoherence time of the order of a microsecond.
Moreover, the upper right plot shows that the sensitivity is increased in the presence of environment.
}
\label{decohere1}
\end{figure}

\begin{figure}[ht]
\centering
\includegraphics[width=8cm,height=6cm]{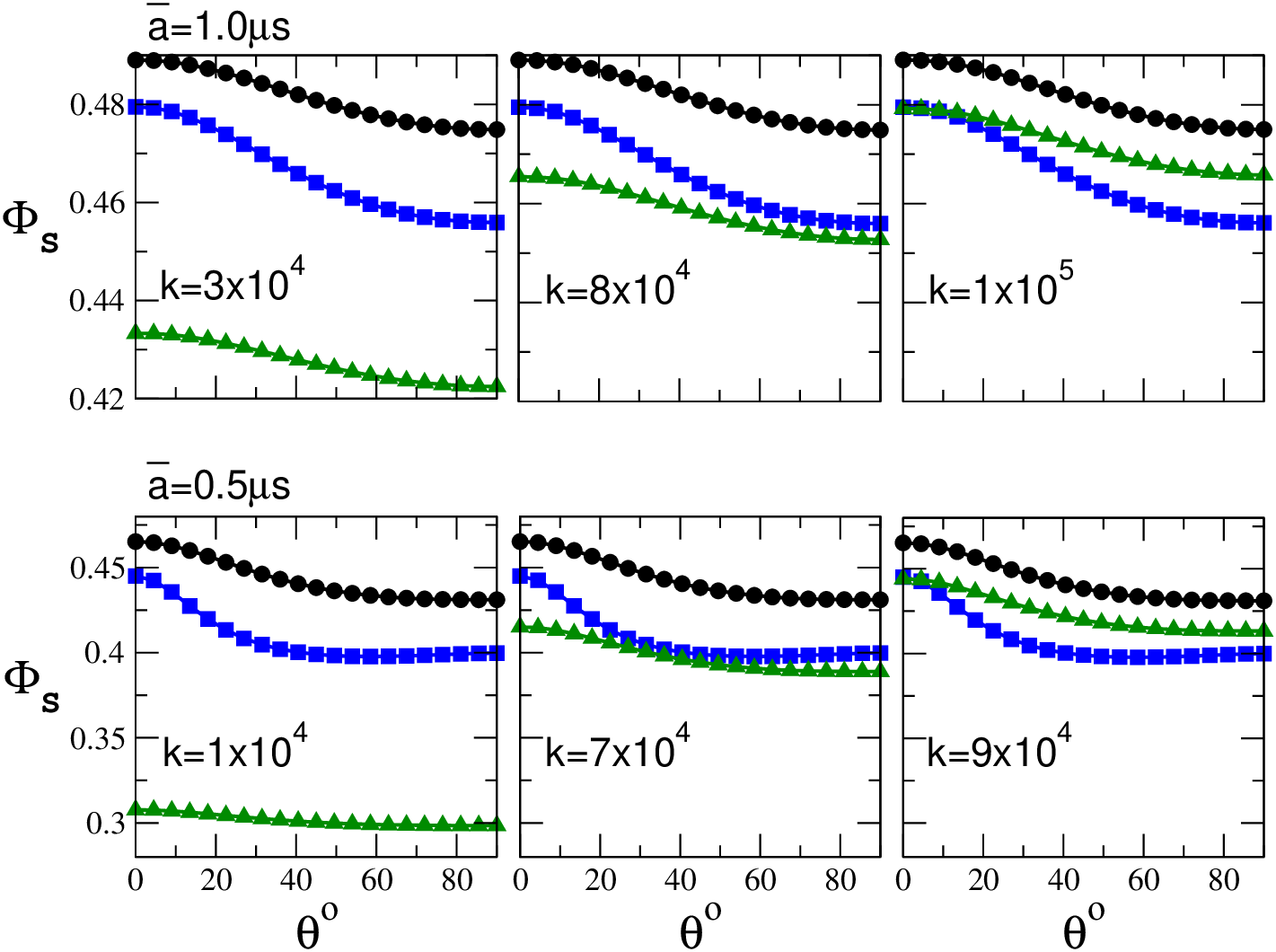}
\caption{(Color online) Life-time estimation of a radical pair using $47 nT$ RF field.
Each plot presents angular dependency of the singlet yield in presence of local geomagnetic field of $47\,\mu T$ $(\bullet)$, in $30\%$ weaker field $(\blacksquare)$, and for the geomagnetic field augmented with the resonant $47 \,nT$ RF field $(\blacktriangle)$.
Here again calculations are performed for two strengths of the hyperfine coupling: slightly smaller than the geomagnetic strength ($\overline{a} = 0.5\, \mu s$, three lower plots) and slight bigger than the geomagnetic strength ($\overline{a} = 1.0\, \mu s$, three upper plots).
On plots from left to right we display the growing value of the recombination rate $k$ used in numerics. Here we estimate the average life-time $11 \mu s$ for $\overline{a} = 1.0\, \mu s$ case and $12.5 \mu s$ for $\overline{a} = 0.5\, \mu s$ case.}
\label{ang_dep2}
\end{figure}

\begin{figure}[ht]
\centering
\includegraphics[width=8cm,height=6cm]{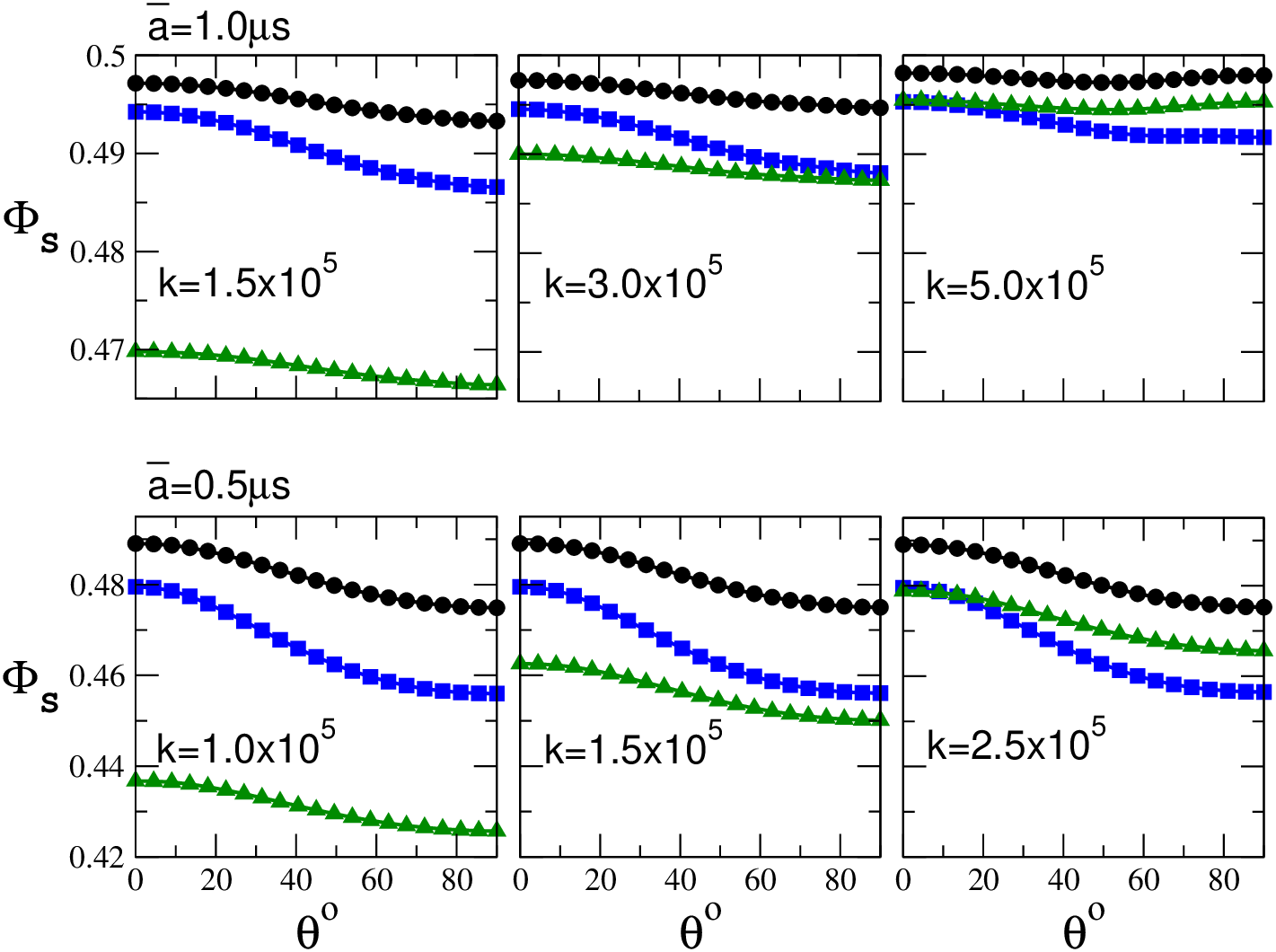}
\caption{(Color online) Life-time estimation of a radical pair for the static field $B_0 = 94 \mu T$ using $150 nT$ RF field.
Each plot presents angular dependency of the singlet yield in presence of artificial magnetic field of $94\,\mu T$ $(\bullet)$, in $30\%$ weaker field $B_0 = 65.8 \mu T (\blacksquare)$, and for the artificial magnetic field augmented with the resonant $150\, nT$ RF field $(\blacktriangle)$.
As usual, calculations are performed for two strengths of the hyperfine coupling: slightly smaller than the geomagnetic strength ($\overline{a} = 0.5\, \mu s$, three lower plots) and slight bigger than the geomagnetic strength ($\overline{a} = 1.0\, \mu s$, three upper plots).
On plots from left to right we display the growing value of the recombination rate $k$ used in numerics. We estimate the average life-time $2.5 \mu s$ for $\overline{a} = 1.0\, \mu s$ case and $5 \mu s$ for $\overline{a} = 0.5\, \mu s$ case.}
\label{ang_dep3}
\end{figure}

\subsection{Resonance of magnetic sensitivity for life-time of about microsecond}

Fig. \ref{decohere1} shows qualitatively similar influence of noise on the sensitivity of the avian compass, as that of Fig. 2 of the main text (values of $k$ are different here). Moreover, the maxima of the resonance for all of these plots correspond to the same noise level.

\subsection{Life-time estimation for RF field of $47\, nT$}

We have also estimated the life-time of RP using much weaker resonant RF field of strength $B_{\mbox{\small rf}} = 47 nT$ ($0.1\%$ strength of the local geomagnetic field). Fig. \ref{ang_dep2} shows that for this case, estimated life-time is $11 \mu s$ for $\overline{a} = 1 \mu s$ and $12.5 \mu s$ for $\overline{a} = 0.5 \mu s$.

\subsection{Life-time estimation for stronger static field}

In Ref. \cite{Ritz2009}, birds were also tested by exposing them to artificially created magnetic field with strength twice as big as the local geomagnetic field. Birds adapted themselves in this stronger magnetic field environment, and hence the artificially created stronger magnetic field is now the effective local geomagnetic field for the birds. The experiment shows that, in this situation, birds are again  disoriented by applying RF field with frequency resonant with the Larmor frequency corresponding to this stronger magnetic field. In addition, according to Ref. \cite{FuncWindow}, the $30\%$ functional window should also exist for this case. Following these, we have performed similar analysis for the stronger $94 \mu T$ artificial magnetic field to estimate the life-time. For this purpose, we use resonant RF field with $B_{\mbox{\small rf}} = 150 nT$, the same RF field which is used in main text. Fig. \ref{ang_dep3} presents results of these calculations which reveal the life-time of $2.5 \mu s$ for $\overline{a}= 1 \mu s$ and $5 \mu s$ for $\overline{a} = 0.5 \mu s$. Both values are again close to those estimated in the main text and show their robustness.

\end{appendix}


\end{document}